# CONTENTS OF PHYSICS RELATED E-PRINT ARCHIVES

*E. R. Prakasan\**, *Anil Kumar\**, *Anil Sagar\**, *Lalit Mohan\**,
*Sanjay Kumar Singh\**, *V. L. Kalyane\** and *Vijai Kumar\**

\*Library and Information Services Division,
Bhabha Atomic Research Centre, Trombay, Mumbai– 400 085, India
E-Mail : prak@magnum.barc.ernet.in; vkalyane@yahoo.com

# CONTENTS OF PHYSICS RELATED E-PRINT ARCHIVES

**Abstract**

The frontiers of physics related e-print archives (1994-2002) at *http://www.arxiv.org/archives/physics* web service are explored from 7770 submissions. No. of e-prints in the six research disciplines besides physics (5390) were: Condensed matter(754), Quantum physics(279), Astrophysics(222), Chemical physics(129), High energy physics – Phenomenology(118), and High energy physics-Theory(100)). By keyword contents following major sub-fields have high frequency: Atomic physics(1258), General physics(1121), Chemical physics(892), Accelerator physics(769), Optics(686), Biological physics(674), and Computational physics(607). Interdomainary co-word cluster analysis revealed higher e-print contents for: Classical physics-General physics(108), Quantum physics-Optics(53), and High energy physics (Phenomenology)-Atomic physics(49). Prominent contributors were B. G. Sidharth (India), V. V. Flambaum (Australia), Antonina N. Fedorova (Russia), and Michael G. Zeitlin (Russia).

**KEYWORDS/DESCRIPTORS:** Scientometrics; Cybermetrics; E-print; Web services; *arxiv.org*; Biological physics; Chemical physics; Astrophysics; Computational physics; Statistical mechanics; Accelerator physics; Free flow of scientific information

# 1 INTRODUCTION

Engineers, scientists, and other researchers have always felt that it is easier to conduct research than publishing it in journals. Hence, the practice of alternative channels (letter writing, poster presentation, demonstration, informal meetings, short notes of work in progress, conferences, technical reports, pre-prints, etc.) for free flow of scientific information.

## 1.1    Non-Peer-Reviewed Pre-Print:

A pre-print is the earliest form of publicly available research [1]. Pre-prints are pre-refereed, pre-publication papers. Pre-prints were a good way of establishing priority. In a fast moving field as soon as research results are available, it is important to get the work into the public domain with a researcher's name attached to it. Publication in journal was usually too slow. Pre-prints on the other hand could be produced quickly and circulated immediately by mail to other researchers in order to establish priority. Pre-prints were a way of soliciting comments on the research so that the paper could be refined for 'formal publication'. On receipt of comments, a paper would then be redrafted for submission to a journal. For the discipline as a whole, pre-prints were a way of reducing the likelihood of unnecessary duplication of parallel research.

However, paper pre-prints were not entirely satisfactory. For a start, they did not halt all disputes over priority. Another key problem had to do with distribution. Distribution was inevitably limited. Only certain institutions received pre-prints; others (including most institutions in underdeveloped countries) were effectively out of the loop. Researchers in these places were at a distinct disadvantage.

**1.2     Non-Peer-Reviewed E-Print:**

The e-print (electronic pre-print) archives were originally designed as a way of automating the paper-based process already in existence. E-print archives are globally accessible open depository of non-peer-reviewed research. Anyone with access to a networked computer can now look at the pre-print literature. Some have seen the e-print archives as 'democratizing' the scholarly communication process. Now the e-print archives services of research output are like newspaper services, with a similar longevity [2].

Scholarly publication forms the intellectual core of any discipline [3]. The provision of free, open, discipline-based access to e-print and re-print are the responses of many scholarly communities to the tactics of the commercial journal publishers who are perceived to profit unreasonably from scholarly work [4-5]. The usefulness will be further enhanced by the implementation of techniques for automatically detecting the occurrence of citations within texts [6], and creating active hyperlinks on that basis [7]. Such techniques have been developed in a series of exploratory projects conducted in a variety of contexts, including the Open Journal project [8], the Open Citation project [9], NEC's ResearchIndex [10], and Ex Libris' SFX framework for dynamic, context-sensitive linking [11-12]. Interoperability among the various emerging "standards" for automated citation-detection and link-creation is one of the primary goals of the Sante Fe Convention [13], developed by the Open Archives Initiative. The citation-detection process itself may be made easier through widespread adoption of a standard format specification such as the Scholarly Link Specification Framework [14].

**1.3   Non-Peer-Reviewed and Peer-Reviewed Physics Related E-Print Archives:**

James Langer, President, American Physical Society, stated that "We need a system for diffusing and archiving scientific information that can grow and change rapidly in size, complexity, and accessibility. Both non-refereed e-print archives and refereed journals would be essential components of that system." [15].

Physicists check the site every day for new information. They post all their papers there, cite references by archives number, use the search engine to find other papers, and need little or no other publication services. Publications on the archives services are instantaneous. It costs the users nothing and is self-organizing. Physicists all over the world can post their research results without being hassled by grumpy editors and referees. In that sense it is far more democratic. They don't have to be part of some inner circle of accepted colleagues to be on the preprint mailing lists and they can find out what's new on the archives just as soon as everyone else does.

Los Alamos National Laboratory (LANL), USA started a multidisciplinary e-print archives services in the early 1990s. It was a brainchild of Paul Ginsparg, [16]. It is currently known as *arxiv.org* (formerly known as Los Alamos XXX service). It processes over 200 new submission per day.

This new system of scientific communication is doing far more than just providing an ultra-effective mode of operation for scientists. It is forcing a complete reevaluation of the role of scholarly journals and, inevitably, an equally thorough reevaluation of the roles of those organizations, as their stated mission is the advancement and diffusion of the knowledge of physics.

The current pattern of usage of *arXiv* looks something like this:

- o A researcher prepares his/her work in one of a number of formats accepted on *arXiv*;
- o The author self-archives by e-mail or FTP or using the submission procedure on the Web and the document gets an *arXiv* document number;
- o Other researchers are making use of it by finding out from the *arXiv* web interface or through e-mail alerting services;
- o Comments of other researchers are getting through e-mails;
- o The author then revises the paper in response to the comments and replaces the original paper with the revised one. The paper may undergo a number of iterations;

- The paper is then submitted to a journal for publication. Some journal publishers now even allow submission in the form of an *arXiv* document number. The referees can refer the paper on *arXiv*;
- As per the referee's comments the paper is either accepted or rejected;
- If rejected, the paper may be submitted to another journal after any necessary revisions. Revised versions may be included on *arXiv*; and
- The process from journal part repeats. The final and revised version is also placed on *arXiv*.

Most preprints are issued with a preprint number assigned by the author's host institutions. This number identifies the paper within the institution and distinguishes it from preprints issued by other institutions. The preprint numbers are not standardized, so it is difficult to group and sort them in a database.

The e-print number assigned by xxx.lanl.gov (the LANL preprint server) provides a standardized common number for preprints that allows the item to be uniquely identified regardless of the institution from which it originated. The e-print number is also useful for citing the work, as well as serving as a common link between databases consisting of bibliographic information and the full text of the article.

LANL's alphanumeric code provides broad subject categorization, year indicator, and accession number. The e-print number is a useful form of identification and serves as a linking point for electronic publications. The SLAC SPIRES database and the Astrophysical Data System (ADS) at Harvard use the e-print number to link their bibliographic (database) records to the full text electronic versions at LANL. Eventually, links could be established using the e-print number (or some other mutually agreed identifier) to track an article throughout its publication process, from inception to final publication, and to reuse the data in future publications [17].

The format of the e-print archives holdings on the web interface includes many links to archives added on the same day, in last five days, different versions of each and every e-print archives, etc.

Studying the service of e-print archives of multidisciplinary nature with a stated mission of the advancement and diffusion of the knowledge of physics is obviously interesting. This paper describes the discipline-wise growth pattern and content analysis of e-print archives for physics and related disciplines available online at *arxiv.org*.

## 2  MATERIALS AND METHODS

A total of 7770 records added to the e-print archives "*http://www.arxiv.org/archives/physics*" on physics related disciplines (including mathematics, non-linear sciences, computational linguistics and neuroscience) during 1994 to 2002 are considered for the present study. Content analysis has been carried out on the three fields: Subj-class; Authors; Journal-ref: for the scientometrics or cybermetrics of archived information.

## 3  RESULTS AND DISCUSSION

### 3.1  Growth of Physics Related E-Print Archives:

Steady growth in the number of physics e-prints was observed recently. Among the total of 7770 e-print archives considered, 75% of them are added during 1999-2002. Research discipline-wise chronological distribution of physics e-print archives are presented in Table 1. Figure 1 depicts the growth pattern of physics e-prints. The growth pattern of six subfields are depicted in Figure 2. All of the subfields have shown a steady and positive growth rate except for Chemical Physics. During the period 1996-1997, there is a rapid growth in the number of e-print archives for 'Quantum Physics' and 'High Energy Physics-Phenomenology'.

| Insert Table 1 here |
| --- |
| Insert Figure 1 here |
| Insert Figure 2 here |

### 3.2 Growth of Physics Subfield in E-Print Archives:

The field 'Subj-class:' in the records has been analysed for their frequency. Table 2 shows high frequency physics subfields in the records as per the classification of LANL. 'Atomic Physics', 'General Physics', 'Chemical Physics', 'Accelerator Physics', 'Optics' 'Biological Physics', and 'Computational Physics', are the major subfields of physics e-print archives. Chronological growth pattern of these seven physics subfields are shown in Figure 3. A steep growth can be observed for the subfields 'Atomic Physics' 'General

Physics' and 'Chemical Physics'. A sudden growth also can be seen for the subfield 'Accelerator Physics' from 1999 to 2000.

> Insert Table 2 here

> Insert Figure 3 here

### 3.3 Interdomainary Clusters of Physics Related E-Prints:

The analysis of the field 'Subj-class:' reveals that many of the physics e-print archives are of intra-disciplinary in nature. Table 3 presents the Interdomainary co-word clusters of subfields in the physics e-print archives. The intra-disciplinary areas of physics 'Classical Physics-General Physics', 'Quantum Physics-Optics', 'High Energy Physics-Atomic Physics' are of high priority to physicists.

> Insert Table 3 here

### 3.4 Prominent Physics Related E-Print Authors:

Physicists are always one step forward in accepting new technological opportunities and challenges. Contributions of individual physicists (Table 4) from India and Australia are high. B. G. Sidharth of Centre for Applicable Mathematics & Computer Sciences, B. M. Birla Science Centre, Hyderabad (India) has contributed 57 items to the e-print archives.

> Insert Table 4 here

### 3.5 Resubmission of Physics Related E-Print to Conference or Journal:

The analysis of the 'Journal-ref:' field has given an idea of how many e-print archives holdings have the information about the source in which these archives are formally published. Among the 7770 item studied 2992 (38.51 %) possess the source details. The analysis of the sources reveals that physicists who are contributing to e-print archives preferentially publish their papers in e-Conferences. *Physical Review Letters*, *Physical Review E*, *Physical Review A*, *Nuclear Instruments and Methods A*, *Journal of Chemical Physics*, *Journal of Physics A,* etc. are some of the most favourite print and/or electronic journals.

Information dissemination behaviour of scientists must synapse with information seeking behaviour of target groups for effective and efficient communication to occur [18]. Librarians have obligatory role to facilitate pre-publication activity in R&D institutions by actively providing information inputs in the knowledge generating system for accelerating knowledge diffusion and dissemination. Librarians must evolve professional management to catalyse the process of bringing together authors and their target readers.

### 4 PROSPECTIVE CONCERNS

Electronic scholarly publishing, and developments of the World Wide Web, are the driving forces behind the implementation of projects for archiving, preservation and provision of access to electronic scientific grey literature [19].

E-print archives are powerful and inexpensive solution for sharing scholarly works with the world, a concept of "Self-archiving" [22]. The e-print archives services are going to run on its own, cost-free and with minimal supervision.

Physics is not the only discipline to have pre-print tradition. Almost all disciplines started archives of their research outputs both for peer-reviewing as well as archival purpose. Hence, database creation, sharing, linking, free exchange, universal accessibility, timeliness, speed, etc. should be concerns for the archival of global intellectual heritage. Logopollution is caused by the production of information that can not reach its potential users. It is dangerous because it makes useful or vital information unavailable. Factors that influence logopollution include message length, assimilation time, message efficiency and availability, message response, merit, and dispersal. Message packaging should assume new forms in order to reach the right recipient [21].

## 5  CONCLUSION

The physics related interdisciplinary and multidisciplinary scientific community have conducted the most innovative and successful experiments in scholarly communication by the use of the *arXiv* server. Researchers use it for disseminating both pre-prints and post-refereed articles. Interestingly, they still wish to have their work accepted by journals, and endorsed by the formal peer-review process, but do not see journals as the only means of dissemination and diffusion of their work. In other words, self-archiving is not seen as a substitute for publishing in peer-reviewed journals, but as a mechanism for accelerating the process of peer-review and improving quality of publication before submitting it to the journal. Past full text e-print archives, which are not published (61.49%) elsewhere, remains the only source for physicists and may complement further research. Present study could reveal some of the current hot areas of interdisciplinary and interdomainary interactions. Many dynamic leaders (researchers, librarians, and R&D managers) are beginning to see the potential of e-print archives.

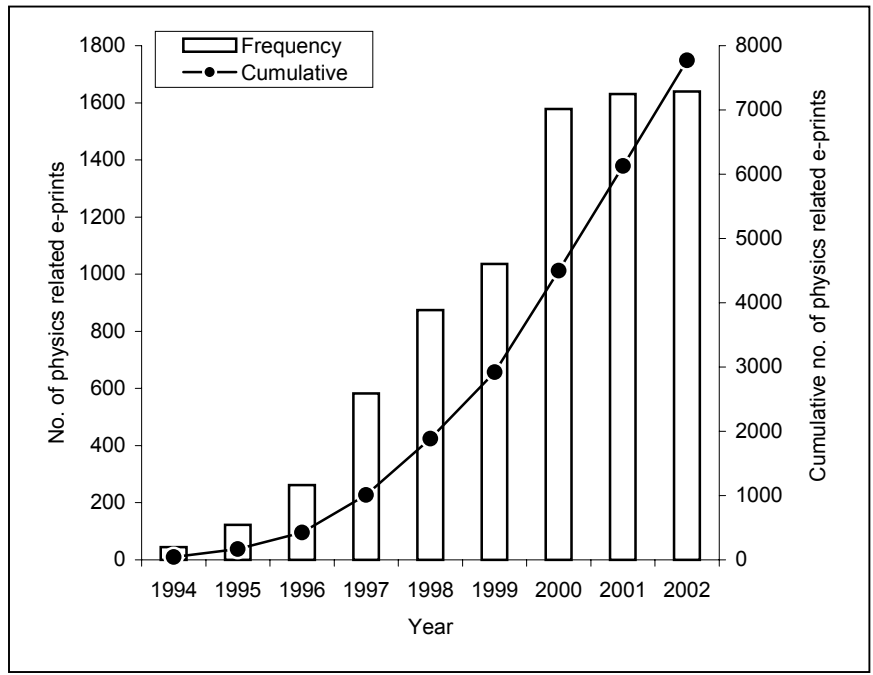

**Figure 1: Year-wise frequency and growth of physics related e-print archives online**
*arxiv.org*

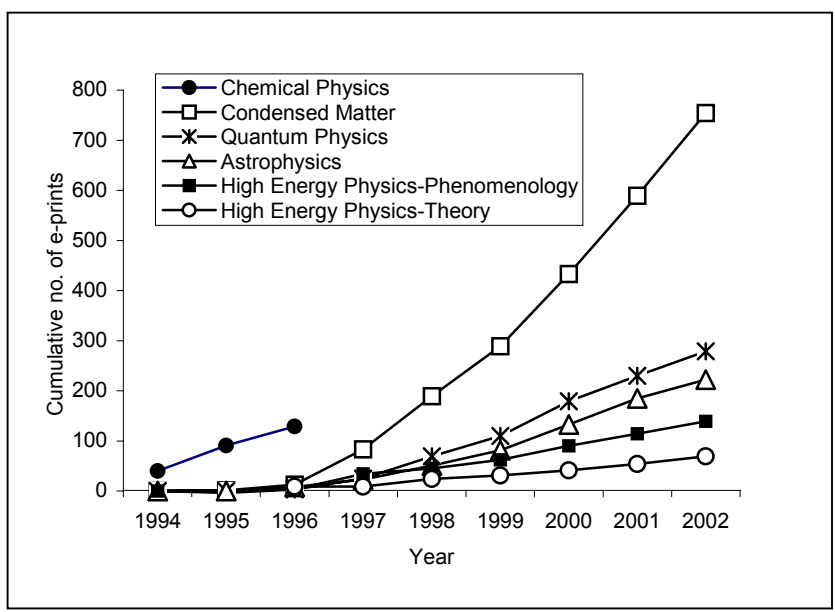

**Figure 2: Growth of six categories of physics related e-print archives online**
*arxiv.org*

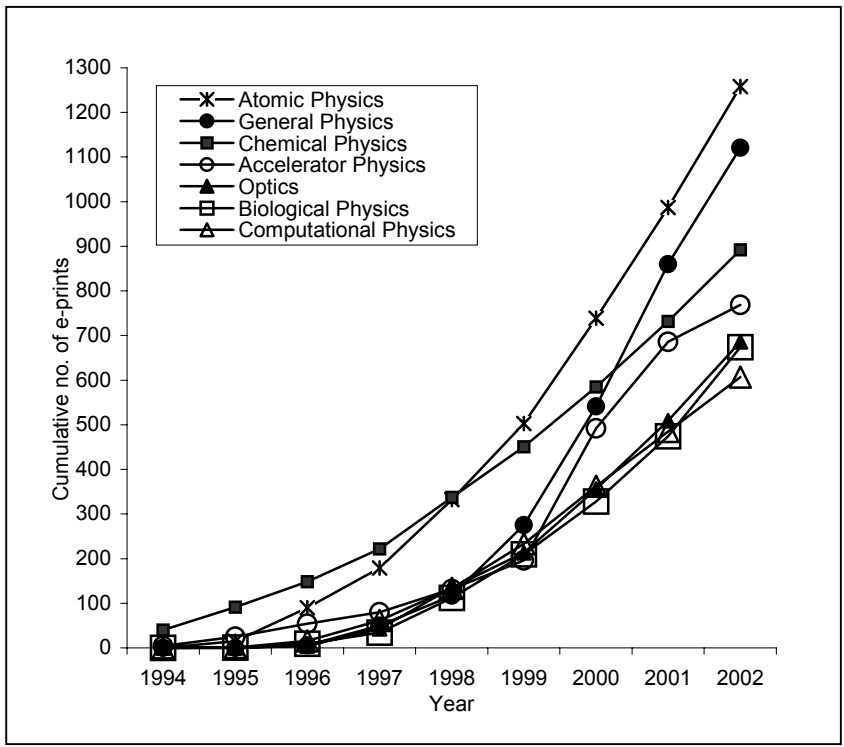

**Figure 3: Growth of prominent subfields in physics related e-print archives online**
*arxiv.org*

**Table 1: Year-wise distribution of physics related e-print archives online *arxiv.org* during 1994-2002**

| Research Discipline | 1994 | 1995 | 1996 | 1997 | 1998 | 1999 | 2000 | 2001 | 2002 | Total |
|---|---|---|---|---|---|---|---|---|---|---|
| Physics | | | 72 | 390 | 589 | 743 | 1144 | 1226 | 1226 | **5390** |
| Condensed Matter | | 1 | 12 | 70 | 106 | 100 | 144 | 156 | 165 | **754** |
| Quantum Physics | | | 3 | 22 | 44 | 41 | 69 | 51 | 49 | **279** |
| Astrophysics | | | 8 | 15 | 28 | 30 | 52 | 52 | 37 | **222** |
| Chemical Physics | 40 | 51 | 38 | | | | | | | **129** |
| High Energy Physics – Phenomenology | | 1 | 2 | 10 | 11 | 17 | 28 | 24 | 25 | **118** |
| High Energy Physics – Theory | | | 9 | 31 | 15 | 7 | 10 | 13 | 15 | **100** |
| General Relativity and Quantum Cosmology | | | 3 | 6 | 12 | 17 | 18 | 15 | 19 | **90** |
| Nonlinear-Chaotic Dynamics | | | | | | | 24 | 31 | 30 | **85** |
| High Energy Physics – Experiment | | 1 | 1 | 2 | 5 | 12 | 36 | 8 | 13 | **78** |
| Nuclear Theory | | | 2 | 9 | 11 | 12 | 15 | 12 | 9 | **70** |
| Atomic Physics | | 15 | 53 | | | | | | | **68** |
| Mathematical Physics | | | | | 5 | 18 | 8 | 10 | 16 | **57** |
| Chaotic Dynamics | | | 2 | 6 | 23 | 18 | | | | **49** |
| Accelerator Physics | 4 | 21 | 22 | | | | | | | **47** |
| Nonlinear-Pattern Formation and Solitons | | | | | | | 12 | 13 | 12 | **37** |
| Plasma Physics | | 16 | 12 | | | | | | | **28** |
| Pattern Formation and Solitons | | | 3 | 3 | 5 | 6 | | | | **17** |
| Quantum Algebra and Topology | | | 7 | 7 | | | | | | **14** |
| Atmospheric and Oceanic Physics | | 8 | 5 | | | | | | | **13** |
| High Energy Physics – Lattice | | | 1 | | 2 | 2 | 5 | | 3 | **13** |
| Nuclear Experiment | | | 1 | 1 | | 2 | 4 | 3 | 2 | **13** |
| Data Analysis, Statistics and Probability | | 8 | 3 | | | | | | | **11** |
| Mathematical Physics-Analysis of PDEs | | | | | 1 | 2 | 2 | 5 | | **10** |
| Adaptation and Self-Organizing Systems | | | 1 | | 5 | 3 | | | | **9** |
| Nonlinear-Adaptation and Self-Organizing Systems | | | | | | | 3 | 1 | 4 | **8** |
| Mathematical Physics-Numerical Analysis | | | | | 1 | | 1 | 3 | 2 | **7** |
| Differential Geometry | | | 1 | 5 | | | | | | **6** |
| Exactly Solvable and Integrable Systems | | | 1 | 2 | 3 | | | | | **6** |
| Computational Physics;Gases | | | | 1 | 2 | 1 | | | | **4** |
| Mathematical Physics-Probability Theory | | | | | | | 1 | 2 | 1 | **4** |
| Computer Science. Computational Complexity | | | | | | | | 2 | 1 | **3** |
| Functional Analysis | | | | | 3 | | | | | **3** |
| Mathematical Physics-Differential Geometry | | | | | 3 | | | | | **3** |
| Nonlinear-Cellular Automata and Lattice Gases | | | | | | | | 1 | 2 | **3** |
| Nonlinear-Exactly Solvable and Integrable Systems | | | | | | | | 1 | 2 | **3** |
| Computer Science. Computer Vision and Pattern Recognition | | | | | | | 1 | 1 | | **2** |
| Computer Science. Distributed, Parallel, and Cluster Computing | | | | | | | | 1 | 1 | **2** |
| Computer Science. Learning; Logic in Computer Science | | | | | | 1 | | | 1 | **2** |
| Computer Science. Neural and Evolutionary Computing | | | | | 1 | | | | 1 | **2** |
| Mathematical Physics-Functional Analysis | | | | | 2 | | | | | **2** |
| Computer Science. Computational Engineering | | | | | | | | | 1 | **1** |
| Computer Science. Computational Geometry | | | | | | | | | 1 | **1** |
| Computer Science. Computers and Society | | | | | | 1 | | | | **1** |
| Computer Science. Symbolic Computation | | | | | | | 1 | | | **1** |
| Mathematical Physics-Combinatorics | | | | | | | | | 1 | **1** |
| Mathematical Physics-Dynamical Systems | | | | | | | 1 | | | **1** |
| Mathematical Physics-Geometric Topology | | | | | | | | | 1 | **1** |
| Mathematical Physics-Representation Theory | | | | | | | 1 | | | **1** |
| Neuroscience-Behavioral Systems | | | | | | | 1 | | | **1** |
| **Total E-prints** | **44** | **122** | **262** | **583** | **874** | **1036** | **1578** | **1631** | **1640** | **7770** |

**Table 2: Frequency of subfields in the physics related e-print archives online *arxiv.org* during 1994-2002**

| Subfield | E-prints | Subfield | E-prints |
|---|---|---|---|
| Atomic Physics | 1258 | Strongly Correlated Electrons | 46 |
| General Physics | 1121 | Exactly Solvable and Integrable Systems | 42 |
| Chemical Physics | 892 | Quantum Algebra | 40 |
| Accelerator Physics | 769 | Dynamical Systems | 35 |
| Optics | 686 | Probability Theory | 34 |
| Biological Physics | 674 | Differential Geometry | 29 |
| Computational Physics | 607 | Numerical Analysis | 29 |
| Statistical Mechanics | 533 | Superconductivity | 29 |
| Fluid Dynamics | 524 | Analysis of PDEs | 22 |
| Plasma Physics | 512 | Cellular Automata and Lattice Gases | 20 |
| Classical Physics | 469 | Mathematical Methods in Physics | 16 |
| Mathematical Physics | 467 | Functional Analysis | 15 |
| Data Analysis, Statistics and Probability | 414 | General Relativity and Coquantum Cosmology | 15 |
| Chaotic Dynamics | 355 | Mesoscopic Systems and Coquantum Hall Effect | 15 |
| Instrumentation and Detectors | 351 | High Energy Physics - Lattice | 13 |
| Soft Condensed Matter | 317 | Nuclear Experiment | 13 |
| Atomic and Molecular Clusters | 286 | Computational Complexity | 11 |
| Quantum Physics | 279 | Algebraic Geometry | 10 |
| Physics Education | 229 | Computational Engineering, Finance, and Science | 10 |
| Astrophysics | 222 | Learning | 10 |
| Atmospheric and Oceanic Physics | 183 | Neural and Evolutionary Computing | 10 |
| Pattern Formation and Solitons | 165 | Combinatorics | 6 |
| Geophysics | 158 | Distributed, Parallel, and Cluster Computing | 6 |
| Condensed Matter | 155 | Artificial Intelligence | 5 |
| Materials Science | 153 | Optimization and Control | 5 |
| Space Physics | 129 | Quantum Algebra and Topology | 5 |
| Disordered Systems and Neural Networks | 122 | Classical Analysis and ODEs | 4 |
| High Energy Physics - Phenomenology | 118 | Computer Vision and Pattern Recognition | 3 |
| History of Physics | 117 | Data Structures and Algorithms | 3 |
| Adaptation and Self-Organizing Systems | 109 | Discrete Mathematics | 3 |
| High Energy Physics - Theory | 100 | Metric Geometry | 3 |
| Popular Physics | 96 | Networking and Internet Architecture | 3 |
| Medical Physics | 92 | Representation Theory | 3 |
| High Energy Physics - Experiment | 78 | Rings and Algebras | 3 |
| General Relativity and Quantum Cosmology | 75 | Spectral Theory | 3 |
| Mesoscopic Systems and Quantum Hall Effect | 74 | Computers and Society | 2 |
| Physics and Society | 71 | Group Theory | 2 |
| Nuclear Theory | 70 | | truncated |

\

**Table 3: Interdomainary co-word clusters observed in the physics related
e-print archives online *arxiv.org* during 1994-2002**

| Co-words | E-prints |
|---|---:|
| Classical Physics - General Physics | 108 |
| Quantum Physics - Optics | 53 |
| High Energy Physics (Phenomenology) - Atomic Physics | 49 |
| Optics - General Physics | 45 |
| Quantum Physics - Atomic Physics | 45 |
| Biological Physics - Chemical Physics | 44 |
| Atomic Physics - Chemical Physics | 39 |
| Biological Physics - Data Analysis, Statistics and Probability | 35 |
| Astrophysics - Plasma Physics | 34 |
| Atomic Physics - Optics | 34 |
| Condensed Matter - Atomic Physics | 32 |
| Atomic Physics - General Physics | 31 |
| Astrophysics - Atomic Physics | 30 |
| Nuclear Theory - Atomic Physics | 30 |
| Condensed Matter - Atomic and Molecular Clusters | 29 |
| Chemical Physics - Computational Physics | 28 |
| Accelerator Physics - Instrumentation and Detectors | 26 |
| Statistical Mechanics - Biological Physics | 26 |
| Atomic and Molecular Clusters - Chemical Physics | 25 |
| Chaotic Dynamics - Fluid Dynamics | 24 |
| Soft Condensed Matter - Chemical Physics | 23 |
| Statistical Mechanics - Chemical Physics | 23 |
| Soft Condensed Matter - Biological Physics | 22 |
| High Energy Physics (Experiment) - Accelerator Physics | 20 |
| | truncated |

**Table: 4 High productive authors ( ≥15 credits ) observed in the physics e-print archives online *arxiv.org* during 1994-2002**

| Author | (Affiliation) | e-prints |
|---|---|---:|
| B. G. Sidharth | (India) | 57 |
| V. V. Flambaum | (Australia) | 46 |
| Antonina N. Fedorova | (Russia) | 37 |
| Michael G. Zeitlin | (Russia) | 37 |
| Holger F. Hofmann | (Japan) | 24 |
| Kirk T. McDonald | (USA) | 24 |
| Sadhan K. Adhikari | (Brazil) | 24 |
| H. C. Rosu | (Mexico) | 22 |
| Lawrence R. Pratt | (USA) | 21 |
| Kikuo Harigaya | (Japan) | 19 |
| R. Jackiw | (USA) | 19 |
| Ulrich H.E. Hansmann | (USA) | 18 |
| E. L. Afraimovich | (Russia) | 18 |
| M. Kibler | (France) | 18 |
| Luis Gonzalez-Mestres | (France) | 18 |
| D. M. Snyder | (-) | 17 |
| Roger Ellman | (USA) | 17 |
| Valery Telnov | (Russia) | 17 |
| Jan ML Martin | (Israle) | 17 |
| P. R. Berman | (USA) | 17 |
| Krzysztof Sacha | (Poland) | 16 |
| Sameen Ahmed Khan | (Mexico) | 16 |
| Uzi Landman | (USA) | 16 |
| Alex Kaivarainen | (Finland) | 15 |
| Alexander A. Vlasov | (Russia) | 15 |
| Constantine Yannouleas | (USA) | 15 |
| D. Andelman | (Israel) | 15 |
| L. Ya Kobelev | (Russia) | 15 |
| R. M. Jones | (USA) | 15 |
| William Bialek | (USA) | 15 |
| E. G. Bessonov | (Russia) | 15 |
| | | truncated |